\documentclass[preprint,11pt,2023]{elsarticle}

\usepackage{amssymb}

\usepackage{geometry}

\journal{Acta Materialia}

\begin{document}

\begin{frontmatter}



\title{Solute Induced Defect Phase Transformations in Mg Grain Boundaries}


\author[1]{Prince Mathews}

\author[1]{Siyuan Zhang}

\author[1]{Christina Scheu}

\author[2]{Rebecca Janisch}

\author[1]{J\"org Neugebauer}

\author[1,3]{Tilmann Hickel}

\affiliation[1]{organization={Max-Planck-Institut für Eisenforschung},
                addressline={Max-Planck-Straße 1}, 
                city={Düsseldorf}, 
                postcode={40237}, 
                country={Germany}}

\affiliation[2]{organization={ICAMS, Ruhr-Universität Bochum},
                addressline={Universitätsstraße 150}, 
                city={Bochum}, 
                postcode={44801}, 
                country={Germany}}

\affiliation[3]{organization={Federal Institute for Materials Research and Testing (BAM)},
                addressline={Richard-Willstätter-Straße 11},
                city={Berlin}, 
                postcode={12489}, 
                country={Germany}}

\begin{abstract}

The study of defect phases is important for designing nanostructured metals and alloys. Grain boundaries (GBs) form one class of defects that directly influence materials properties, such as deformability and strength. At the same time, alloying can introduce GB phase transformations and can therewith alter mechanical performance. In this work, the defect phases of a $\Sigma$7 $(12\bar{3}0)$ [0001] 21.78$^\circ$ symmetric tilt GB in hcp Mg are investigated. Ab-initio simulations as a function of stress and temperature (using quasi-harmonic approximation) are performed, and different types of phase transformations are revealed. To this end, the influence of the chemical degree of freedom on the defect phases is studied for the example of Ga addition, using an efficient screening approach that combines empirical potentials and accurate ab-initio calculations. By exploiting the concept of defect phase diagrams, a phase transformation from the T to the A structural type and as well as  a systematic transition of the segregation site preference is revealed. The results qualitatively agree well with experimental observations from scanning transition electron microscopy. The underlying physical mechanisms have an impact on grain-boundary engineering in metallic alloys. 

\end{abstract}




\end{frontmatter}


\section{Introduction}\label{sec1}
As a potential material for automotive and aerospace applications, various Mg alloys are being explored to improve mechanical properties like yield strength, ductility and fracture toughness. Grain boundaries (GBs)  are defects that are known to have a strong influence on the mechanical properties of the material \cite{hallpetch}. GBs can have different atomic configurations when subjected to different conditions (stress, temperature or chemical composition) which have been referred to as defect phases \cite{korte2022defect} or complexions \cite{cantwell2014gb, dillon2007complexion}. Although these defect phases may only exist locally within the environment of the defect, they still follow the same thermodynamic principles as bulk phases, and can transition from one defect phase to another. Such defect phase transitions can occur at pure GBs where the structural arrangement of the atoms at the GB transforms due to a change in the intensive state variables, as has been reported for Cu $\Sigma$5 GBs with change in temperature \cite{frolov2013Cu}. Many different GBs can exist in a material and with alloying, different concentration of solutes are expected to segregate to different GBs. Furthermore, different solutes can preferentially segregate to specific GBs \cite{zn_in_fe, o2018grain}. Computational studies are particularly helpful for GBs that exhibit segregation of low concentrations of solutes. Unlike solute-rich GBs, where accurate computational studies are difficult to perform as finding the energetically stable configurations and positions of solute atoms are non-trivial at high solute concentrations due to solute-solute interactions becoming dominant. These interactions can cause abrupt changes in the GB character, majorly influencing and determining not only the stability of defect phases causing defect phase transitions at higher solute concentrations, but also the properties of the material like in the case of Ga diffusion to Al GBs \cite{ga_in_al_gb_2, ga_in_al_gb_3}. Similar observations have already been reported by Sickafus et al. \cite{sickafus1987grain} in Fe-Au alloys where a structural transformation was reported at solute-rich small angle [001] twist grain boundaries. Therefore to design tailored materials, studying the chemical degree of freedom in GBs to build defect phase diagrams (DPDs) can be a challenging but rewarding task. \par

A vast number of studies have been performed on surface segregation of Ga in semiconductors and for thermoelectric applications. Ga alloying to GeTe has been studied where Ga atoms segregated to GBs\cite{Ga_study_1}. For engineering applications, Ga in Al alloys have been studied extensively as it induces embrittlement as a result of segregation to Al GBs \cite{ga_in_al_gb}. Ga is very commonly used in transmission electron microscope (TEM) sample preparation processes where focused ion beam (FIB) is used with a Ga$^+$ ion source. This often leads to the formation or modification of defects in the sample due to implantation of Ga atoms onto the sample surface. The implanted Ga atoms may diffuse into the sample, as is known that Ga$^+$ FIB milling of Al, and Mg can lead to segregation of Ga into GBs of Al \cite{ga_fib_al} and Mg \cite{tem_method}, while in Cu, Cu-Ga precipitates have been observed at the sample surface \cite{ga_fib_cu}. This makes the study of the effect of Ga inclusion to materials interesting as Ga decoration of specific defects can be captured, having the potential to reveal defect phase transitions within these individual defects which could be difficult to capture starting from an alloy containing Ga. \par

Mg-Ga alloys have not been as extensively researched as compared to Mg alloys containing cheaper metals like Al, Ca and Zn. However, existing literature shows that Ga improves mechanical performance when alloyed to Mg, in contrast to the effect it has in Al. In the work by Kub$\acute{\rm a}$sek et al., Mg-Ga alloys were suggested as the most prospective for biodegradable implants over other Mg-based binary alloys considered due to better strength, hardness and corrosion resistance \cite{Mg_Ga_kubasek}. While other studies on engineering applications of Mg-Ga alloys reported Ga to activate pyramidal $<$c+a$>$ slip, improve yield strength, ultimate strength (both tensile and compressive), better work hardening ability and ductility as compared to pure Mg \cite{Mg_Ga_huang, Mg_Ga_liu, Mg_Ga_kubasek16, Mg_Ga_huang22}. The strengthening effect of Ga in Mg has been mainly attributed to solid solution strengthening, while the formation of Mg$_5$Ga$_2$ intermetallic phase also plays a role. Nevertheless, only a few studies have been performed on the segregation behavior of Ga to crystal defects, with the concentration of Ga restricted to very dilute limits. \par

In general, various studies have been conducted on GBs and segregation of solutes to GBs in Mg \cite{Mg_seg_1,Mg_seg_2, Mg_seg_3, Mg_seg_4, Mg_seg_5, Mg_seg_6, Mg_seg_7, Mg_seg_8, Mg_seg_9}. Due to direct consequences on plasticity, a lot of studies, both experimental and theoretical, have been performed on $\{10\bar{1}2\}$ and $\{10\bar{1}1\}$ twins in Mg including segregation of various solutes within the dilute limit. In a study by Nie et al. \cite{nie_Mg}, periodic segregation of Zn and Gd atoms to contraction and extension sites respectively at the twin boundary was reported. High-throughput first principles studies on segregation of different elements to $\{10\bar{1}2\}$ and $\{10\bar{1}1\}$ Mg twin boundaries show the segregation of solute atoms smaller in size (like Ga) and solute atoms larger in size than Mg to contraction and extension lattice sites respectively \cite{Mg_twin_seg_1, Mg_twin_seg_2, nie2020microstructure}. Similar observations were reported in the study by Huber et al. \cite{huber_Mg} on Mg $\Sigma$7 GB, where atoms smaller in size than Mg like Ag, Zn and Al show preferential segregation to contraction sites and atoms larger in size than Mg prefer extension sites.

Taking into account the existing literature and the limited supercell size to perform DFT calculations, the Mg $\Sigma$7 $(12\bar{3}0)$ [0001] 21.78$^{\circ}$ symmetric tilt GB is chosen as a suitable candidate for this work and the Mg-Ga system is chosen as an example system to study defect phase transitions and construction of defect phase diagram. Two structural types (defect phases) are known for this $\Sigma$7 GB, the A and the T type \cite{wang1997tilt}. However, both structure types have only been experimentally observed in ZnO \cite{sato2007atomic}. In this article, the two defect phases for the pure $\Sigma$7 GB are investigated using simulation and experiments. Based on experimental evidence, the influence of segregation of Ga in the very dilute limit on the GB is explored, along with a few  technologically relevant solutes of Al, Ca and Gd. Furthermore, the segregation of Ga going beyond the dilute limit is systematically investigated in order to look for defect phase transformations. Eventually, the DPD for the $\Sigma$7 GB as a function of chemical potential of Ga is constructed.

\section{Methodology}\label{sec2}
\subsection{Computational Details}\label{DFT_calculations}
The DFT calculations have been performed using the Vienna Ab initio Simulation Package (VASP) \cite{vasp_1, vasp_2}. The ionic cores and valence electron interaction are described by applying the projected augmented wave method \cite{vasp_3} and the exchange-correlation effects are described by the generalized gradient approximation (GGA) applying the Perdew-Burke-Ernzerhof (PBE) form of parameterization. Convergence tests were performed on Mg bulk and the following parameters were selected based on energy convergence $<$ 1 meV: plane wave cutoff energy of 550 eV, k-point spacing of 0.012 \AA$^{-1}$ along all directions (based on the Monkhorst-Pack scheme \cite{monkhorst_pack}) and a smearing width of 0.15 eV (using the Methfessel-Paxton \cite{methfessel_paxton} smearing scheme). The molecular statics (MS) simulations have been carried out using Large-scale Atomic/Molecular Massively Parallel Simulator (LAMMPS) \cite{lammps}.  \par

Phonon calculations to calculate vibrational contribution to the free energy employing quasi harmonic approximation have been performed using the Phonopy package \cite{phonopy_1, phonopy_2} with VASP applying the small displacement method. Starting from the 0 K equilibrium structure as calculated from DFT, nine different volumes have been considered for supercells of GBs and bulk structures with a q-mesh of $30 \times 30 \times 30$ grid points. For the case of GBs, complete relaxation of the supercells have been performed before the phonons are calculated.\par

The lattice constants for hcp Mg have been determined by calculating energy-volume curves (DFT) for different \textit{c/a} ratios and comparing the lowest energy for the optimum \textit{a} lattice parameter at each of the \textit{c/a} ratios. The computed lattice constants are \textit{a} = 3.192 \AA\ and \textit{c/a} = 1.624 for hcp Mg and are in very good agreement with the experimentally determined lattice constants of \textit{a} = 3.203 \AA\ and \textit{c/a} = 1.623 \cite{exp_Mg_lc}. As the construction of symmetric tilt GBs with tilt axis [0001] are not dependent on the value of the lattice parameter \textit{c}, the $\Sigma$7 GB can be constructed by joining two perfect grains each rotated by half the angle of the GB and removing atoms that overlap onto the other grain. The selection of this cutting plane to remove the overlapping atoms determines the structure type of the $\Sigma$7 GB. Then, the atoms close to each other are merged to get the A and T structure types. Due to geometry, there are two GBs in the supercells and the distance between the two GBs is approximately 4 nm.

After the two GBs are constructed, the strain state of the supercells is optimized to determine the equilibrium structure. Atomic relaxation in the supercells are performed while subjecting them to different strains (both positive and negative) along the direction normal to the GB, while preserving the original lattice constants in the part of the supercells that are bulk like. There are 158 atoms for the A type and 162 atoms for the T type supercell, and the dimensions of the GB unit cell are 5.185\AA\ in the [0001] direction with 8.444\AA\ in the [2$\bar{1}\bar{1}$0] direction.

Using the equilibrium structures, the grain boundary energy (GBE) is calculated using the following equation:
\begin{equation} \label{GBE_eq}
    \gamma_{\rm GB} = \frac{E_{\rm GB}-E_{\rm bulk}}{2A_{\rm GB}}, 
\end{equation}
where E$_{\rm{GB}}$, E$_{\rm{bulk}}$ and A$_{\rm{GB}}$ are the energy of the supercell containing the GBs, the rescaled energy of hcp Mg bulk based on the number of atoms in the supercell containing the GBs and the area of cross-section of the GB in the supercell. A correction factor of two is included in the denominator as a penalty for the two GBs in the supercell.

The segregation energies for different solute concentrations were calculated by replacing Mg atoms by solute atoms at different sites close to one of the GBs in the supercell, following which strain optimization is performed. The segregation energy of the solute atoms is calculated using:
\begin{equation} \label{seg_eq}
E_{\rm seg} = \Big[E_{\rm{GB}} \left( \rm{Mg}_{\textit{N-m}} \textit{X}_{\textit{m}} \right) - E_{\rm{GB}}\left(\rm{Mg}_{\textit{N}}\right)\Big] - \Big[E_{\rm{bulk}}\left(\rm{Mg}_{\textit{M}-1}\textit{X}\right) - E_{\rm{bulk}}\left(\rm{Mg}_{\textit{M}}\right)\Big], 
\end{equation}
where \textit{E}$_{\rm{GB}}$(Mg$_{N-m}$X$_{m}$) is the energy of the GB supercell with the solute atoms, \textit{E}$_{\rm{GB}}$(Mg$_{N}$) is the energy of the GB supercell without solute atoms, \textit{E}$_{\rm{bulk}}$(Mg$_{M-1}$X) is the energy of the bulk structure containing one solute atom and \textit{E}$_{\rm{bulk}}$(Mg$_{M}$) is the energy of the bulk structure without solute atoms.\par

To build the DPD, the formation energies of the phases are calculated using the following equation:
\begin{equation} \label{form_en}
E_{\rm{f}} = \frac{E_{\rm{sup}} - M\mu_{\rm{Mg}} - N\mu_{\rm{Ga}}}{2A_{\rm{GB}}},
\end{equation}
where \textit{E}$_{\rm{sup}}$ is the energy of the supercell containing the defect phase, \textit{$\mu$} represents the chemical potential, and \textit{M} and \textit{N} are the number of Mg and Ga atoms respectively, in the supercell.

The chemical potential for solid solution of Ga in Mg is the energy change due to introduction of a Ga atom in Mg bulk supercell, and is calculated as follows:
\begin{equation} \label{mu_ss}
    \mu^{\rm ss}_{\rm Ga} = E_{\rm Mg_{N-1}Ga_1} - (N-1)\mu_{\rm Mg}
\end{equation}
where $E_{\rm Mg_{N-1}Ga_1}$ is the energy of Mg bulk supercell containing 1 Ga atom, $N$ is the total number of atoms in the supercell and  $\mu_{\rm Mg}$ is the chemical potential of Mg. Using this chemical potential, the concentration of Ga can be determined by:
\begin{equation} \label{cGa}
    \mu_{\rm Ga} = \mu^{\rm ss}_{\rm Ga} + k_{\rm B}T \ln(c_{\rm Ga})
\end{equation}
where k$_{\rm B}$ is the Boltzmann constant, $T$ is the temperature and $c_{\rm Ga}$ is the concentration of Ga. However, this equation may only be used when Ga exists in Mg bulk as solid solution.

\subsection{Experimental conditions}\label{TEM}
Atomic resolution micrographs on the T type and A type GB structures were taken using a Titan Themis scanning transmission electron microscope (STEM) operated at 300 kV. Detailed experimental conditions can be found in the report \cite{tem_method}. Note that the focused ion beam (FIB) preparation using Ga beam would result to Ga segregation at the GB and hence an A type structure. The T type structure was obtained in a sample prepared by FIB using Xe beam.

\section{Results and Discussion}\label{sec3}

\begin{figure}[!ht]%
\centering
\includegraphics[width=0.3\textwidth]{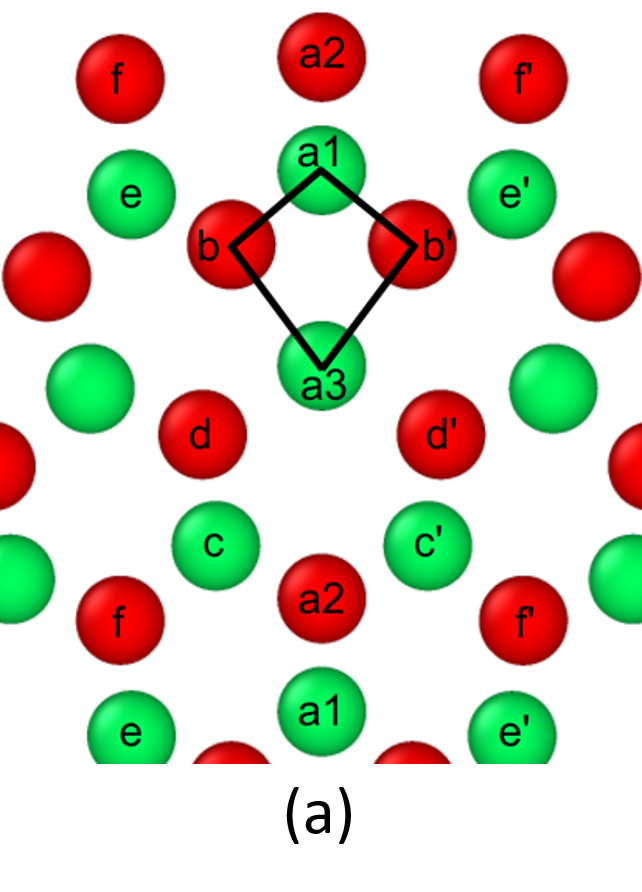}
\includegraphics[width=0.3\textwidth, height=6.2cm]{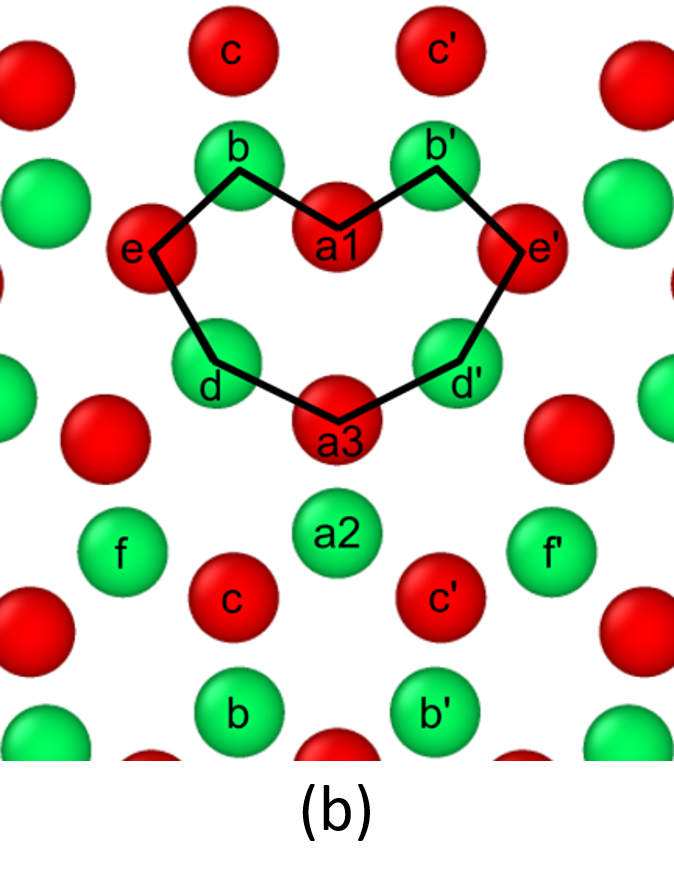}\par
\includegraphics[width=0.3\textwidth]{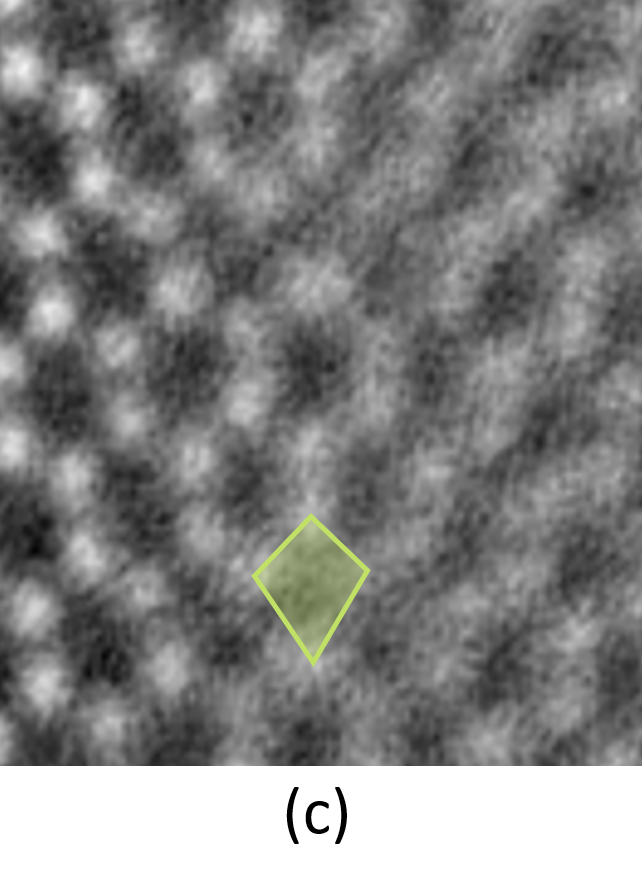}
\includegraphics[width=0.3\textwidth, height=6.3cm]{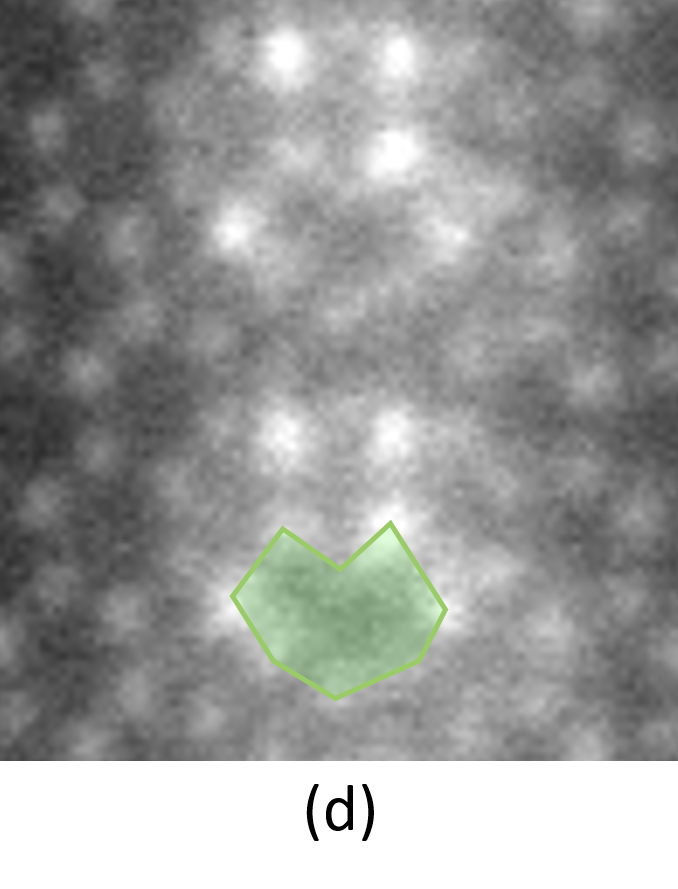}
\caption{Motifs of (a) T type and (b) A type $\Sigma$7 GB along with the considered segregation sites at the GB. The experimental structures observed for the $\Sigma$7 GB in (c) pure Mg with the T type structure and (d) with Ga segregation with the A type structure.} \label{motifs}
\end{figure}

The motifs of the two structure types of $\Sigma$7 GB, marked in Fig.~\ref{motifs}a (T type) and b (A type), are followed by perfect hexagons in both types which along with the motifs constitute the GB unit cell. The atom colors in Fig.~\ref{motifs}a and b represent the different positions of atoms along the [0001] direction. The notation of the non-equivalent sites within the vicinity of the GB are taken from the work of Huber et al. \cite{huber_Mg}, where the atoms at the GB plane are denoted as \textit{a1}, \textit{a2} and \textit{a3} and moving in the direction normal to the GB plane with increasing distance the atoms are denoted in the alphabetical order from \textit{b} and \textit{b'} to \textit{f} and \textit{f'} on either sides of the GB respectively. Sites \textit{b} and \textit{e} in both types have a lower Voronoi volume compared to that in bulk and are known as the contraction or compressive sites, while sites \textit{a3} and \textit{d} have a higher Voronoi volume and are known as the extension sites. \par

Fig.~\ref{motifs}c and d show the STEM images for the clean Mg $\Sigma$7 GB and for the Ga decorated GB due to Ga$^+$ ion FIB preparation. For better visualization, the structure types have been marked in the images. In general, a good agreement is seen between the experimentally obtained GB and the ones predicted in simulation (Fig.~\ref{motifs}a and b). In the undecorated GB, the T type is observed whereas for the Ga decorated GB the structure type changes to A type, indicating a defect phase transition. This motivates the study of stability of both the defect phases for the clean GB.

\subsection{Stability of Types}\label{stability_types}

\begin{figure}[!b]%
\centering
\includegraphics[width=\textwidth]{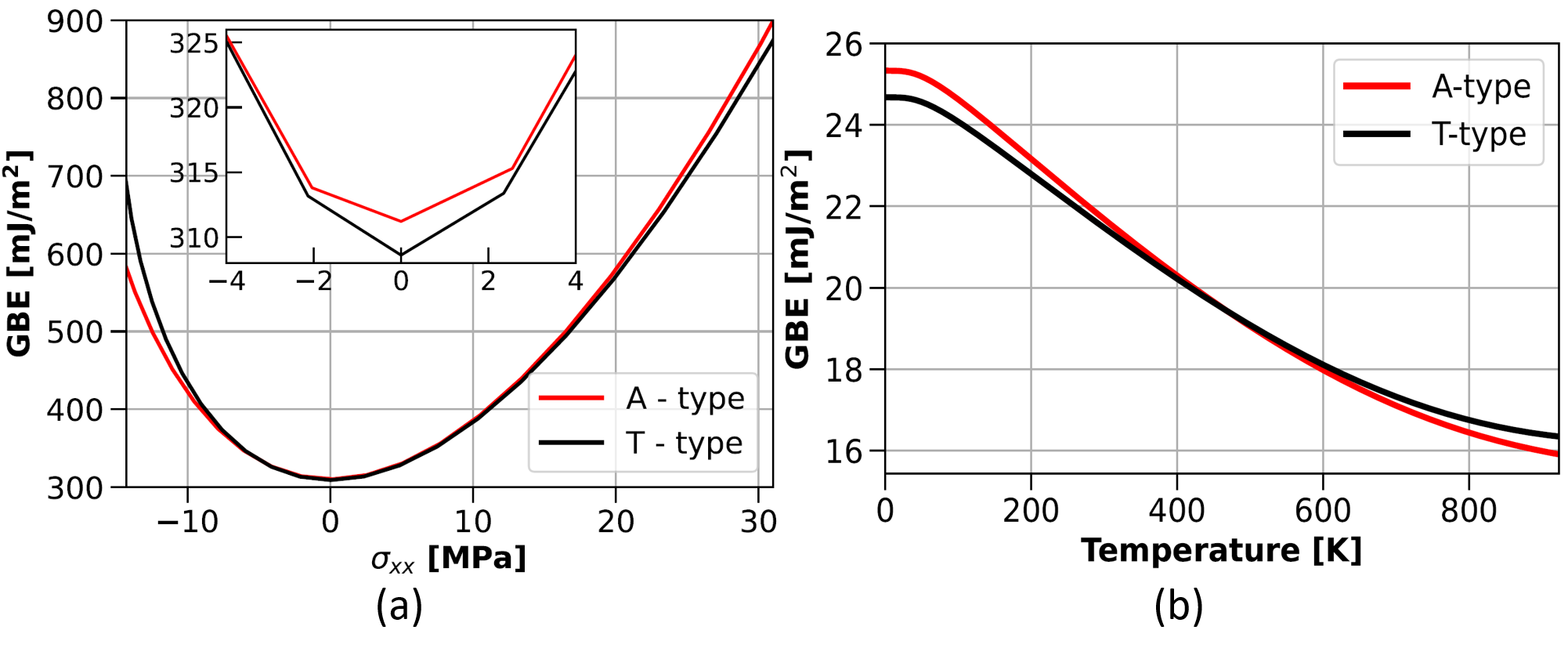}
\caption{GBE as a function of (a) stress in the direction normal to the GB ($x$ direction) and (b) temperature. The inset plot in (a) shows the GBEs close to $\sigma_{xx}=0$ MPa. GBEs for the A and T type have been plotted in red and black, respectively, in both plots.} \label{stability}
\end{figure}

The GBEs, using the optimized supercells (as mentioned in section~\ref{DFT_calculations}), at the lowest energy strain states for the two types applying Eq.~\ref{GBE_eq} are calculated as 311.2 and 308.6 mJ/m$^{2}$ for the A and T types respectively. The excess volumes for the equilibrium states as calculated for the A type is 0.27 \AA\ whereas for the T type it is 0.22 \AA\, with the change in excess volume being 18\%. The GBEs are plotted as a function of the stress (induced due to the applied strain) in the direction normal to the GB in Fig.~\ref{stability}a. The inset plot shows the GBEs close to 0 stress (equilibrium), where the T type is more stable. The T type is still the stable structure type in the tensile (positive) stress region, whereas in the compressive (negative) stress region, the A type becomes stable beyond 8 MPa. The transition from T to A type as a function of stress is therefore a 'defect phase transition' \cite{korte2022defect}.

Additionally, the stability of the structure types have also been investigated as a function of temperature employing quasi-harmonic approximation to calculate the free energy for the supercells containing the GBs and Mg bulk. The temperature dependent GBEs have been plotted in Fig.~\ref{stability}b also taking into account the change in volume due to thermal expansion. At 0 K, there is a strong reduction in grain boundary energy due to complete relaxation of the supercells and consideration of quantum mechanical effects, which suggests that the $\Sigma$7 GB should be easily observable in experiments as the GBE is inversely proportional to the GB area. At low temperatures, the T type is still observed to be the stable defect phase. With increase in temperature, the meta-stable A type shows a stronger decrease in free energy compared to the T type and becomes the more stable defect phase at higher temperatures, with the transition temperature around 450 K for this first order defect phase transition. The proximity of the GBEs around the transition temperature strongly suggests that experimentally both types should be observable between 400 and 550 K. Also considering, the experimentally seen structure of T type for pure Mg GB at room temperature (Fig.~\ref{motifs}c), the calculated temperature dependent GBEs show a good correlation between experiments and theoretical calculations.

The discontinuities observed in GBEs as a function of stress and temperature suggest that the two configurations are not micro-states but indeed two defect phases. Following the experimental evidence of segregation of Ga to the A type of $\Sigma$7 GB, in the TEM image in Fig.~\ref{motifs}, the change in defect phase chemistry due to segregation is explored to understand the effect of Ga inclusion to the $\Sigma$ 7 GBs. To this end, only the sites marked in Fig.~\ref{motifs} for the two types are considered for this segregation study, as considering sites further away from the GB are in the bulk region.  

\subsection{Solute Segregation}

\subsubsection{Dilute limit}

\begin{figure}[!b]%
\centering
\includegraphics[width=\textwidth]{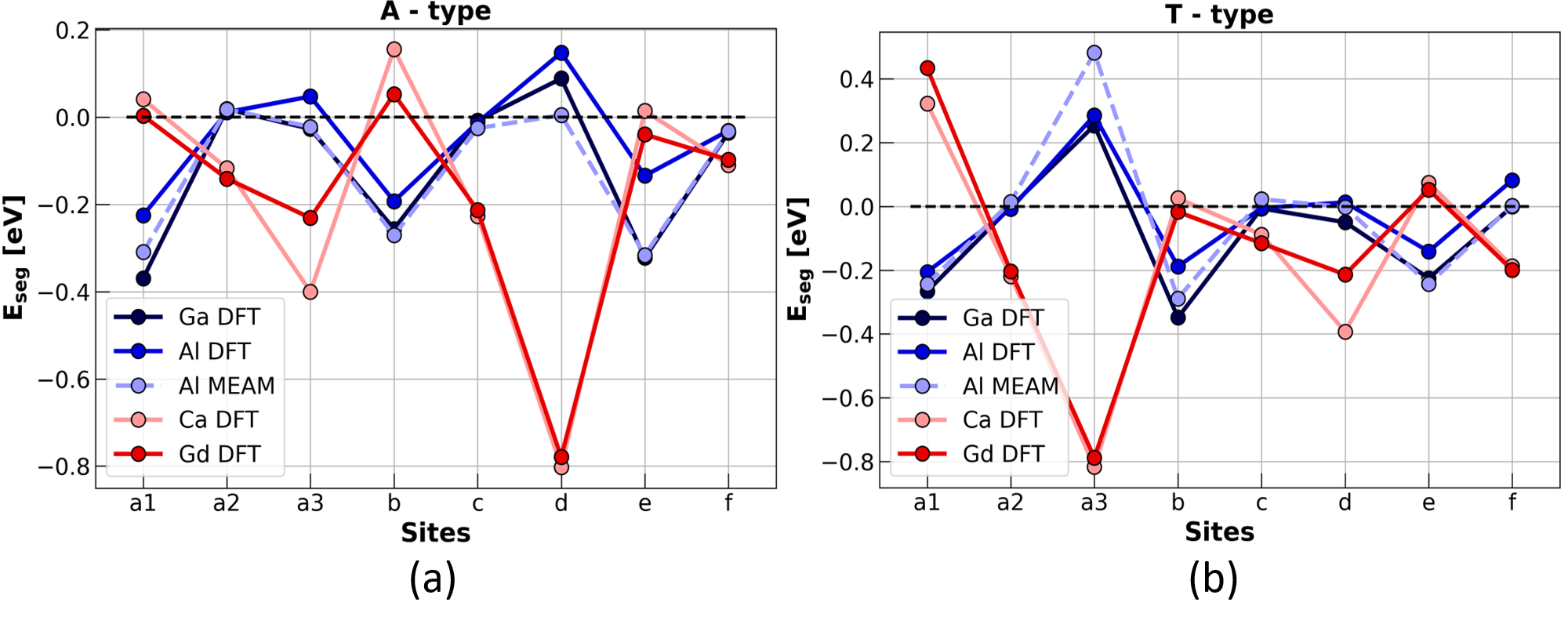}
\caption{Single-atom segregation energy for different solutes at sites in the (a) A and (b) T type GB structures. The segregation energies calculated by DFT have been labeled as ``DFT'', while the segregation energies calculated by molecular statics have been labeled as ``MEAM'' in both plots.}\label{1Ga_segregation}
\end{figure}

In order to understand the influence of Ga segregation to the GBs, the segregation energy of 1 Ga atom per structural unit has been calculated using Eq.~\ref{seg_eq} at all the sites considered in the two types and is shown in Fig.~\ref{1Ga_segregation}. A negative segregation energy corresponds to an energy gain and a positive segregation energy corresponds to an energy loss due to segregation. Along with Ga, the segregation profile for a few relevant solutes has also been calculated. Al is similar in size to a Ga atom and both atoms are smaller than Mg. In contrast to Al and Ga the segregation profile for Ca and Gd has been calculated as they are larger in size compared to Ga, Mg and Al. The computed segregation profiles for Al and Ca in this work are in good agreement to the ones reported by Huber et al. \cite{huber_Mg}. The preference of segregation of the Ga and Al atoms to the sites \textit{a1}, \textit{b} and \textit{e} is seen within both, the A and the T types. The most favourable site for segregation of Ga and Al is site \textit{a1} in the A type. Contrary to existing literature, the site \textit{a1} is not a contraction site and has a slightly larger Voronoi volume compared to that of an Mg atom in bulk, however, the presence of the solute at this site results in reduction of Voronoi volume due to minimisation of strain energy, making it a favorable segregation site for atoms smaller in size than Mg. Meanwhile, Ca and Gd exhibit very favorable segregation to the $\Sigma$7 GB and the preference of segregation to sites \textit{a3} and \textit{d} are seen, however, site \textit{a3} in the T type is the most favorable. The solute atoms also induce transition between types, Ga and Al at sites \textit{b} and \textit{e} in the T type transform the GB from T to the A type whereas Ca and Gd at site \textit{d} in the A type transform the GB to the T type. Therefore, from the induced transitions and the stable sites in the respective structure types, atoms smaller in size than Mg, Ga and Al stabilize the A type while atoms larger in size than Mg, Ca and Gd stabilize the T type GB. These findings from DFT align with the experimental observation of A type in the TEM image showing Ga segregation in Fig.~\ref{motifs}d and confirm the stabilization of A type with Ga. In Mg, similar stabilisation of defect phase with segregation of solutes is also seen in $\{10\bar{1}1\}$ and $\{10\bar{1}2\}$ twins, where Y stabilizes the reflection twin over the glide twin \cite{pei_2017_Mg}. Hence, here the first chemistry induced defect phase transition is reported in Mg, where the T type GB transforms to A type with Ga segregation to the GB.\par

\subsubsection{Beyond the Dilute Limit}\label{high_Ga}

In search for more chemistry induced defect phase transformations, more Ga atoms are introduced at the GB systematically going beyond the dilute limit up to 6 Ga atoms, and energetically favorable atomic configurations at each resulting solute concentration are looked into. There are about a combined 2100 unique possible ways to arrange 2 to 6 Ga atoms at the considered sites in either GBs. As performing DFT simulations for all of these configurations, including optimization of energies, would take tremendous computational time and energy, MS is employed to pre-screen all configurations (refer to section~\ref{pre-screen}). No binary empirical potential for Mg-Ga is available, hence, based on similar segregation behaviour of Al and Ga seen in Fig.~\ref{1Ga_segregation}, Al is chosen as a proxy for Ga to perform the screening procedure. A binary modified embedded atom method (MEAM) potential for Mg-Al \cite{kim_Mg_Al_pot} was identified and the segregation profile of Al (1 atom at the GB) calculated using the MEAM potential is plotted in Fig.~\ref{1Ga_segregation}a and b (dashed line). Surprisingly, the segregation profile of Al using the potential matches to that of Ga more than Al calculated from DFT. Therefore, after the pre-screening for all configurations is performed using the MEAM potential, upscaling to the energetically favorable arrangements for all solute concentrations is done. The energies calculated by DFT for these upscaled configurations for 2 and 6 Ga atoms at the GB are shown in Fig.~\ref{pre-screen-fig}. 

\begin{figure}[!t]%
\centering
\includegraphics[width=0.74\textwidth]{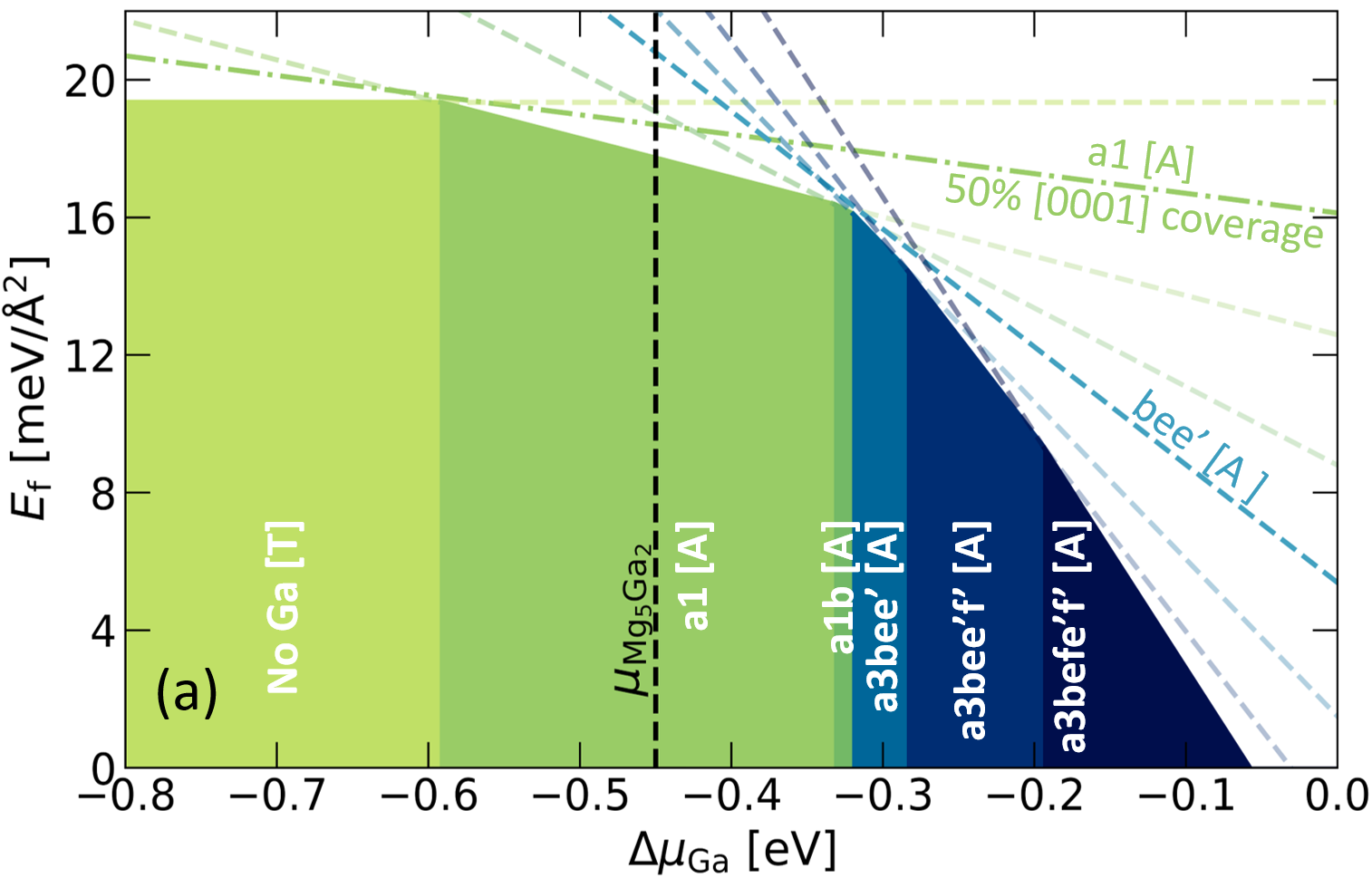}
\includegraphics[width=0.74\textwidth]{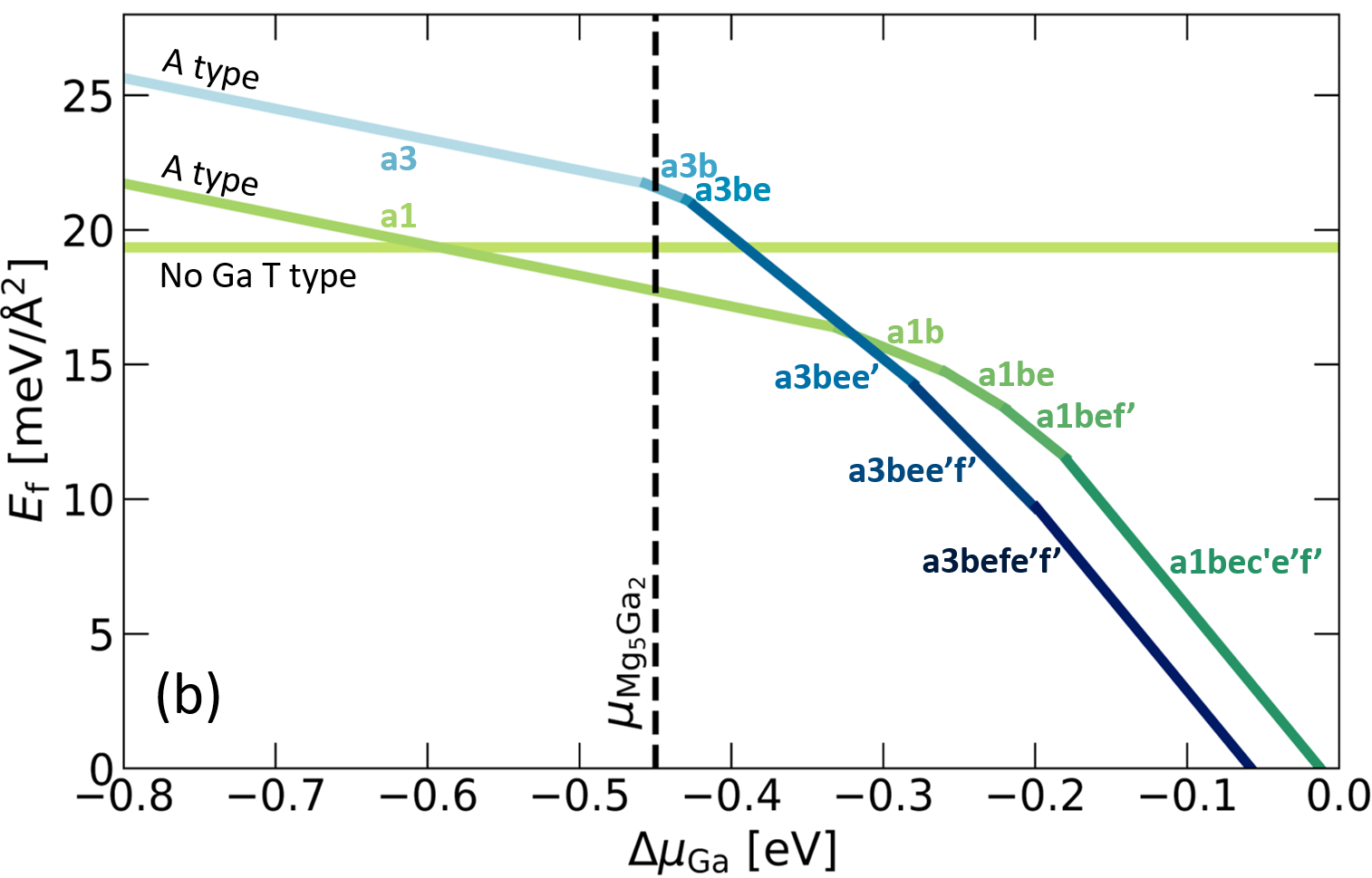}
\caption{(a) Defect phase diagram for the Mg $\Sigma$7 GB plotted as a function of $\Delta\mu_{\rm Ga} = \mu_{\rm Ga} - \mu^{\rm bulk}_{\rm Ga}$. The structure type of the GB is mentioned in square brackets after the sites occupied by Ga atoms, and the number of sites mentioned represent the number of Ga atoms at the GB. (b) Energy of formation for the lowest energy configurations with successive filling of Ga atoms at sites at the GB starting from site \textit{a1} (green) and \textit{a3} (blue). The sites mentioned for each configuration represent the number of Ga sites at the GB. $\Delta\mu_{\rm Ga} = 0$ eV represents the chemical potential of Ga bulk and the vertical dashed line at $\Delta\mu_{\rm Ga} = -0.45$ eV represents the chemical potential of the Mg$_5$Ga$_2$ phase in both plots.}\label{DPD}
\end{figure}

Using the configurations with the lowest energy at each solute concentration, the DPD for Mg $\Sigma$7 GB has been constructed as shown in Fig.~\ref{DPD}a. The energies of formation, calculated using Eq.~\ref{form_en}, have been plotted against $\Delta\mu_{\rm Ga} = \mu_{\rm Ga} - \mu^{\rm bulk}_{\rm Ga}$, where $\mu^{\rm bulk}_{\rm Ga}$ is the chemical potential of Ga in Ga bulk and $\mu_{\rm Ga}$ is considered as a free parameter. 
For the undecorated GB, the T type is stable as was already discussed in section~\ref{stability_types}. At $\Delta \mu_{\rm Ga}$ $>= -0.592$ eV (corresponds to a concentration of $<$1e-05 at.\% Ga at 300~K as calculated using Eq.~\ref{cGa}), the A type becomes stable with the segregation of Ga to the \textit{a1} site of the GB . For Ga atoms at the GB, replacing a Mg atom at a site in the simulation supercell by Ga physically means introducing a column of atoms in the GB, as the cell length in the direction along the GB for [0001] CSL GBs is the lattice parameter \textit{c}. Therefore, considering supercell size limitations with DFT, the formation energies for GBs have been calculated with 50\% partial column coverage in the [0001] direction and have been plotted in the DPD. As partial occupation is less stable compared to 100\% coverage, partial occupations for the other configurations are not considered. \par 

The A type remains stable with further addition of Ga. At $\Delta \mu_{\rm Ga} = -0.332$ eV, Ga segregation to \textit{a1 b} at the GB becomes stable, after which at $\Delta \mu_{\rm Ga} = -0.32$ eV direct jump to 4 atom occupation of sites (\textit{a3 b e e'}) at the GB is seen. However, the site preference also shifts from \textit{a1} to \textit{a3}, following which there is a systematic filling of sites with the same \textit{a3} configuration upto 6 Ga atoms. For 3 Ga atoms at the GB, the most stable configuration is \textit{b e e'}, but as the \textit{a3 b e e'} configuration is energetically more stable, a region for \textit{b e e'} doesn't appear in the DPD. To better understand the shift of site preference, the configurations with successive filling of Ga at the GB, starting from sites \textit{a1} and \textit{a3} in the A type GB have been shown in Fig.~\ref{DPD}b. The crossover from \textit{a1} to \textit{a3} configurations occurs at $-0.32$ eV, and at higher chemical potentials, \textit{a1} configurations become less stable. The change in preference of segregation to an extension site in pure GB is unexpected considering 1 atom Ga segregation profile in Fig.~\ref{1Ga_segregation}. This shows the importance of the consideration of solute-solute interactions in segregation studies. Therefore, the \textit{a3} site only becomes a favorable site for segregation at a higher solute concentration, due to solute-solute interactions. Therefore, this transition at a higher solute concentration is the second chemistry induced defect phase transformation. \par

It is noteworthy that in the region of stability of 6 Ga atom segregation, the formation energy goes below 0 meV/\AA$^{2}$ at very high chemical potentials. This formation is exothermic in nature, and the energy gained through such exothermic processes is used by the system to decrease the chemical potential in order to acquire enough energy to cross the barrier to get to a state with lower chemical potential and with further crossing of barriers eventually achieving equilibrium. The solubility of Ga in Mg at low temperatures is very low, at 500 K the solubility limit is $\approx$1 at.\%. Therefore at low temperatures, the formation of the closest Mg-rich intermetallic phase is expected to form at even lower concentrations, which is Mg$_5$Ga$_2$. The vertical dashed line labeled $\mu_{\rm Mg_5Ga_2}$ in Fig.~\ref{DPD}a and b is the difference in chemical potential of Ga in the intermetallic phase Mg$_{\rm 5}$Ga$_{\rm 2}$ and in bulk. This chemical potential represents the equilibrium state where Ga solid solution in bulk and precipitates of Mg$_5$Ga$_2$ coexist. However, for the GB under consideration, the equilibrium state would change to decoration of the GB with Ga at site \textit{a1} along with the formation of the intermetallic phase. An important factor that is decisive for the formation of this phase is the nucleation barrier (at $\Delta\mu_{\rm Ga} = 0.156$ eV), that needs to be overcome, either by applying energy externally, for example by mechanical work, or internally through exothermic processes. The high energy barrier of 0.606 eV could possibly lead to formation of precipitates not seen at 0.01 Ga at.\% (corresponding to $\Delta\mu_{\rm Ga} = -0.45$ eV) but instead with a higher concentration of Ga, the precipitates would first need to nucleate and the system would then return to the equilibrium state. \par

\section{Conclusions}\label{sec4}
The Mg-Ga alloy system has been explored as an example system to construct a defect phase diagram and to search for possible defect phase transformations. The defect structure investigated for this purpose is the $\Sigma$7 $(12\bar{3}0)$ [0001] 21.78$^\circ$ symmetric tilt GB, which is known to have two defect phases, A and T type, depending on different atomic configurations at the GB. 
An efficient approach to pre-screen configurations at high solute concentrations has been suggested in this work, to explore the defect phase chemistry space using DFT.
The main highlights from this work can be summarized as follows:
\begin{itemize}
    \item The stability of the two phases have been studied as a function of stress (normal to the GB) and temperature. The T type is the equilibrium structure at 0K. Two first order defect phase transitions are reported, namely the stabilization of the A type by compressive stress and the stabilization of the A type structure at temperatures above 450 K.

    \item The segregation of Ga, Al, Ca, and Gd to the two structure types in the $\Sigma$7 GB has been studied. Atoms smaller in size than Mg, Al and Ga stabilize the A type, whereas atoms larger than Mg stabilize the T type. However, Ca and Gd exhibit a higher binding energy as compared to Al and Ga. The first chemistry induced defect phase transformation is the stabilization of the A type with Ga segregation to the GB, this transformation is also experimentally captured with STEM.
    
    \item From the segregation energies calculated by DFT for the selected configurations, the lowest energy configurations for 2 to 6 Ga atoms at the $\Sigma$7 GB have been chosen and used for the construction of the  DPD as a function of the Ga chemical potential.
    
    \item A second chemistry induced defect phase transformation is revealed in the DPD, where the segregation preference changes systematically from \textit{a1} to \textit{a3} site. After the transition, successive filling of sites preserving the 4 atom \textit{a3 b e e'} configuration is seen. The \textit{a3} site only becomes a favorable site for segregation at solute-rich conditions. 
    
    \item Considering the solubility limit, the segregation of more than 1 atom per structural unit to the GB only happens, when the solid solution is oversaturated. Therefore, the precipitation of the nearest Mg-rich intermetallic phase is expected, and the equilibrium state is identified as the segregation of 1 atom at the GB along with precipitates of Mg$_5$Ga$_2$. However, the precipitates only appear in the system once the system overcomes the energy barrier of nucleation. 
    
\end{itemize}


\section*{Acknowledgement}
The authors would like to acknowledge Deutsche Forschungsgemeinschaft (DFG) for providing financial support through projects C05 and B01 within the Collaborative Research Centre SFB 1394 “Structural and Chemical Atomic Complexity - From Defect Phase Diagrams to Materials Properties” (Project ID 409476157).

\appendix

\section{Pre Screening} \label{pre-screen}

\begin{figure}[!h]%
\centering
\includegraphics[width=0.75\textwidth]{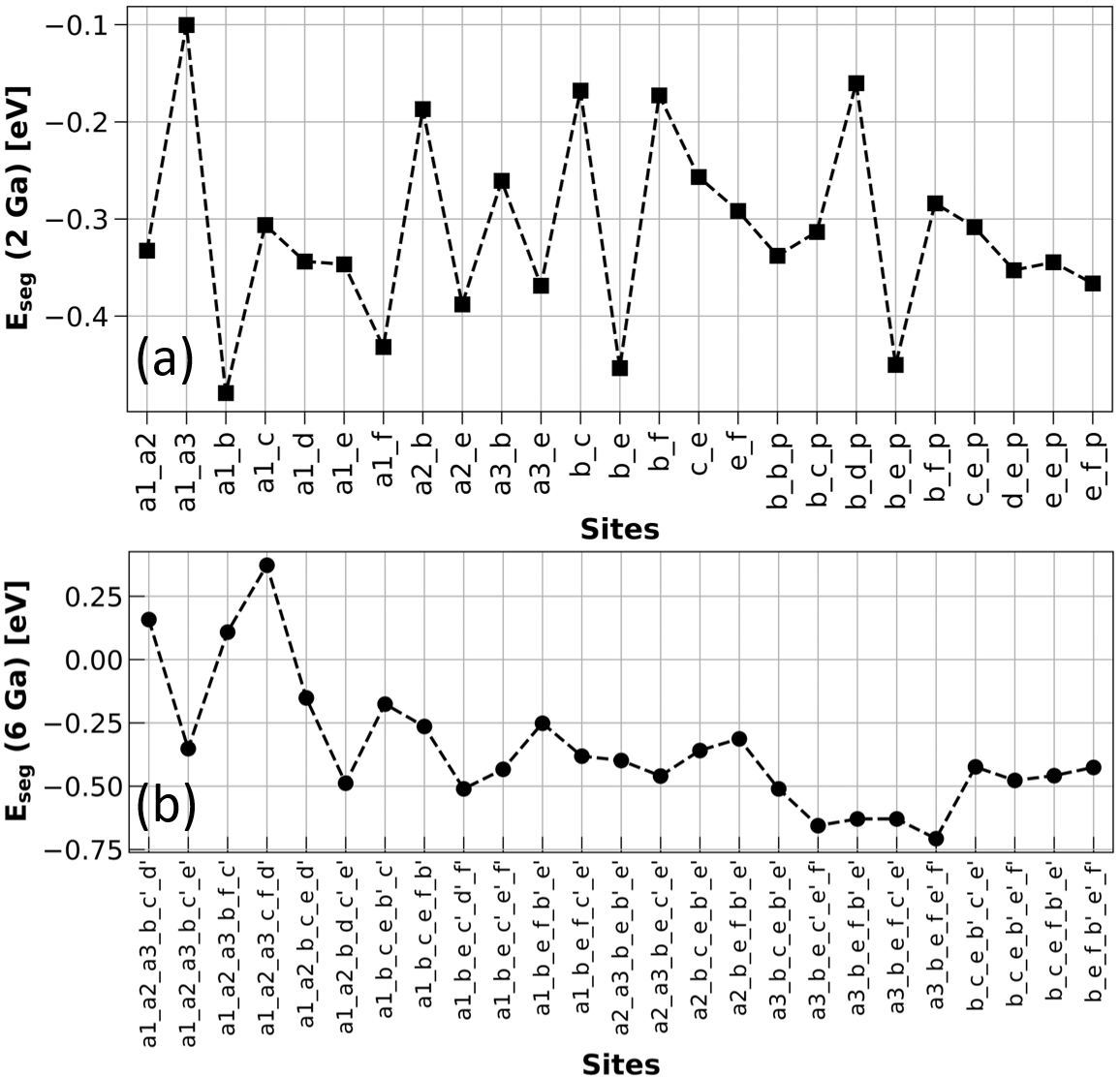}
\caption{Segregation energies calculated for the upscaled configurations for (a) 2 Ga atoms at the GB, (b) 6 Ga atoms at the GB. The labels in the x-axis refer to the sites occupied by Ga.}\label{pre-screen-fig}
\end{figure}

To select configurations to be upscaled to DFT calculations, the segregation energies are calculated for all the different configurations using the Mg-Al MEAM potential as mentioned in section~\ref{high_Ga}. The configurations for 2 and 3 Ga atoms that show the lowest segregation energies are then selected. For the case of more than 4 Ga atom configurations, an additional criteria for selecting configurations for upscaling is having a clear A or T type structure at the GB of the structure type at the GB. This is done as the GB structure may change substantially at times and result in structures that cannot be identified as the GB under consideration. Out of the 43 different configurations for 2 atoms and 878 configurations for 6 atoms, 25 configurations were selected. Fig.~\ref{pre-screen-fig} shows the segregation energies calculated using DFT for these upscaled configurations belonging to 2 and 6 atoms.

\newpage




\bibliographystyle{elsarticle-num} 
\bibliography{bibdatabase.bib}

\begin{thebibliography}{10}
\expandafter\ifx\csname url\endcsname\relax
  \def\url#1{\texttt{#1}}\fi
\expandafter\ifx\csname urlprefix\endcsname\relax\def\urlprefix{URL }\fi
\expandafter\ifx\csname href\endcsname\relax
  \def\href#1#2{#2} \def\path#1{#1}\fi

\bibitem{hallpetch}
N.~Hansen, Hall--petch relation and boundary strengthening, Scripta materialia
  51~(8) (2004) 801--806.

\bibitem{korte2022defect}
S.~Korte-Kerzel, T.~Hickel, L.~Huber, D.~Raabe, S.~Sandl{\"o}bes-Haut,
  M.~Todorova, J.~Neugebauer, Defect phases--thermodynamics and impact on
  material properties, International Materials Reviews 67~(1) (2022) 89--117.

\bibitem{cantwell2014gb}
P.~R. Cantwell, M.~Tang, S.~J. Dillon, J.~Luo, G.~S. Rohrer, M.~P. Harmer,
  Grain boundary complexions, Acta Materialia 62 (2014) 1--48.

\bibitem{dillon2007complexion}
S.~J. Dillon, M.~Tang, W.~C. Carter, M.~P. Harmer, Complexion: A new concept
  for kinetic engineering in materials science, Acta Materialia 55~(18) (2007)
  6208--6218.

\bibitem{frolov2013Cu}
T.~Frolov, D.~L. Olmsted, M.~Asta, Y.~Mishin, Structural phase transformations
  in metallic grain boundaries, Nature communications 4~(1) (2013) 1899.

\bibitem{zn_in_fe}
G.~Jung, I.~S. Woo, D.~W. Suh, S.-J. Kim, Liquid zn assisted embrittlement of
  advanced high strength steels with different microstructures, Metals and
  Materials International 22 (2016) 187--195.

\bibitem{o2018grain}
C.~O’Brien, C.~Barr, P.~Price, K.~Hattar, S.~Foiles, Grain boundary phase
  transformations in ptau and relevance to thermal stabilization of bulk
  nanocrystalline metals, Journal of Materials Science 53~(4) (2018)
  2911--2927.

\bibitem{ga_in_al_gb_2}
Y.~Zhang, G.-H. Lu, M.~Kohyama, T.~Wang, Investigating the effects of a ga
  layer on an al grain boundary by a first-principles computational tensile
  test, Modelling and Simulation in Materials Science and Engineering 17~(1)
  (2008) 015003.

\bibitem{ga_in_al_gb_3}
Y.~Zhang, G.-H. Lu, T.~Wang, S.~Deng, X.~Shu, M.~Kohyama, R.~Yamamoto,
  First-principles study of the effects of segregated ga on an al grain
  boundary, Journal of Physics: Condensed Matter 18~(22) (2006) 5121.

\bibitem{sickafus1987grain}
K.~E. Sickafus, S.~Sass, Grain boundary structural transformations induced by
  solute segregation, Acta Metallurgica 35~(1) (1987) 69--79.

\bibitem{Ga_study_1}
C.~Zhang, G.~Yan, Y.~Wang, X.~Wu, L.~Hu, F.~Liu, W.~Ao,
  O.~Cojocaru-Mir{\'e}din, M.~Wuttig, G.~J. Snyder, et~al., Grain boundary
  complexions enable a simultaneous optimization of electron and phonon
  transport leading to high-performance gete thermoelectric devices, Advanced
  Energy Materials 13~(3) (2023) 2203361.

\bibitem{ga_in_al_gb}
W.~Ludwig, D.~Bellet, Penetration of liquid gallium into the grain boundaries
  of aluminium: a synchrotron radiation microtomographic investigation,
  Materials Science and Engineering: A 281~(1-2) (2000) 198--203.

\bibitem{ga_fib_al}
K.~A. Unocic, M.~J. Mills, G.~Daehn, Effect of gallium focused ion beam milling
  on preparation of aluminium thin foils, Journal of microscopy 240~(3) (2010)
  227--238.

\bibitem{tem_method}
S.~Zhang, Z.~Xie, P.~Keuter, S.~Ahmad, L.~Abdellaoui, X.~Zhou, N.~Cautaerts,
  B.~Breitbach, S.~Aliramaji, S.~Korte-Kerzel, et~al., Atomistic structures of
  $<$0001$>$ tilt grain boundaries in a textured mg thin film, Nanoscale
  14~(48) (2022) 18192--18199.

\bibitem{ga_fib_cu}
J.~Marien, J.~Plitzko, R.~Spolenak, R.~Keller, J.~Mayer, Quantitative electron
  spectroscopic imaging studies of microelectronic metallization layers.,
  Journal of microscopy 194~(1) (1999) 71--78.

\bibitem{Mg_Ga_kubasek}
J.~Kub{\'a}sek, D.~Vojt{\v{e}}ch, J.~Lipov, T.~Ruml, Structure, mechanical
  properties, corrosion behavior and cytotoxicity of biodegradable mg--x (x=
  sn, ga, in) alloys, Materials Science and Engineering: C 33~(4) (2013)
  2421--2432.

\bibitem{Mg_Ga_huang}
W.~Huang, J.~Chen, H.~Yan, W.~Xia, Ga alloying for fabricating magnesium alloy
  sheet with uniform microstructure and excellent performance, Materials
  Letters 304 (2021) 130607.

\bibitem{Mg_Ga_liu}
H.~Liu, G.~Qi, Y.~Ma, H.~Hao, F.~Jia, S.~Ji, H.~Zhang, X.~Zhang, Microstructure
  and mechanical property of mg--2.0 ga alloys, Materials Science and
  Engineering: A 526~(1-2) (2009) 7--10.

\bibitem{Mg_Ga_kubasek16}
J.~Kub{\'a}sek, D.~Vojt{\v{e}}ch, D.~Dvorsk{\`y}, Structural and mechanical
  study on mg--xlm (x= 0--5 wt.\%, lm= sn, ga) alloys, International Journal of
  Materials Research 107~(5) (2016) 459--471.

\bibitem{Mg_Ga_huang22}
W.~Huang, J.~Chen, H.~Yan, W.~Xia, B.~Su, High plasticity mechanism of high
  strain rate rolled mg-ga alloy sheets, Journal of Materials Science \&
  Technology 101 (2022) 187--198.

\bibitem{Mg_seg_1}
J.~Hadorn, T.~Sasaki, T.~Nakata, T.~Ohkubo, S.~Kamado, K.~Hono,
  \href{https://www.sciencedirect.com/science/article/pii/S1359646214003352}{Solute
  clustering and grain boundary segregation in extruded dilute mg–gd alloys},
  Scripta Materialia 93 (2014) 28--31.
\newblock \href
  {https://doi.org/https://doi.org/10.1016/j.scriptamat.2014.08.022}
  {\path{doi:https://doi.org/10.1016/j.scriptamat.2014.08.022}}.
\newline\urlprefix\url{https://www.sciencedirect.com/science/article/pii/S1359646214003352}

\bibitem{Mg_seg_2}
J.~D. Robson, S.~J. Haigh, B.~Davis, D.~Griffiths, Grain boundary segregation
  of rare-earth elements in magnesium alloys, Metallurgical and Materials
  Transactions A 47 (2016) 522--530.

\bibitem{Mg_seg_3}
J.~D. Robson, Effect of rare-earth additions on the texture of wrought
  magnesium alloys: the role of grain boundary segregation, Metallurgical and
  Materials Transactions A 45 (2014) 3205--3212.

\bibitem{Mg_seg_4}
X.~Cai, B.~Sun, Y.~Liu, N.~Zhang, J.~Zhang, H.~Yu, J.~Huang, Q.~Peng, T.~Shen,
  Selection of grain-boundary segregation elements for achieving stable and
  strong nanocrystalline mg, Materials Science and Engineering: A 717 (2018)
  144--153.

\bibitem{Mg_seg_5}
J.~Zuo, T.~Nakata, C.~Xu, Y.~Xia, H.~Shi, X.~Wang, G.~Tang, W.~Gan, E.~Maawad,
  G.~Fan, et~al., Effect of grain boundary segregation on microstructure and
  mechanical properties of ultra-fine grained mg--al--ca--mn alloy wires,
  Materials Science and Engineering: A 848 (2022) 143423.

\bibitem{Mg_seg_6}
T.~Nakata, Z.~Li, T.~Sasaki, K.~Hono, S.~Kamado, Role of grain boundary
  segregation on microstructural development in basal-textured mg-al-zn alloy
  sheet, Scripta Materialia 218 (2022) 114828.

\bibitem{Mg_seg_7}
F.~Mouhib, R.~Pei, B.~Erol, F.~Sheng, S.~Korte-Kerzel, T.~Al-Samman,
  Synergistic effects of solutes on active deformation modes, grain boundary
  segregation and texture evolution in mg-gd-zn alloys, Materials Science and
  Engineering: A 847 (2022) 143348.

\bibitem{Mg_seg_8}
Z.~Zhang, J.~Zhang, J.~Xie, S.~Liu, W.~Fu, R.~Wu, Developing a mg alloy with
  ultrahigh room temperature ductility via grain boundary segregation and
  activation of non-basal slips, International Journal of Plasticity (2023)
  103548.

\bibitem{Mg_seg_9}
R.~Pei, Z.~Xie, S.~Korte-Kerzel, J.~Gu{\'e}nol{\'e}, T.~Al-Samman, Atomistic
  origin of the anisotropic grain boundary segregation in a mg-mn-nd alloy,
  arXiv preprint arXiv:2201.02884 (2022).

\bibitem{nie_Mg}
J.~F. Nie, Y.~Zhu, J.~Liu, X.-Y. Fang, Periodic segregation of solute atoms in
  fully coherent twin boundaries, Science 340~(6135) (2013) 957--960.

\bibitem{Mg_twin_seg_1}
J.~Zhang, Y.~Dou, Y.~Zheng, Twin-boundary segregation energies and
  solute-diffusion activation enthalpies in mg-based binary systems: a
  first-principles study, Scripta Materialia 80 (2014) 17--20.

\bibitem{Mg_twin_seg_2}
Z.~Pei, R.~Li, J.-F. Nie, J.~R. Morris, First-principles study of the solute
  segregation in twin boundaries in mg and possible descriptors for mechanical
  properties, Materials \& Design 165 (2019) 107574.

\bibitem{nie2020microstructure}
J.~Nie, K.~Shin, Z.~Zeng, Microstructure, deformation, and property of wrought
  magnesium alloys, Metallurgical and Materials Transactions A 51 (2020)
  6045--6109.

\bibitem{huber_Mg}
L.~Huber, J.~Rottler, M.~Militzer, Atomistic simulations of the interaction of
  alloying elements with grain boundaries in mg, Acta materialia 80 (2014)
  194--204.

\bibitem{wang1997tilt}
Y.~Wang, H.~Ye, On the tilt grain boundaries in hcp ti with [0001] orientation,
  Philosophical Magazine A 75~(1) (1997) 261--272.

\bibitem{sato2007atomic}
Y.~Sato, T.~Yamamoto, Y.~Ikuhara, Atomic structures and electrical properties
  of zno grain boundaries, Journal of the American Ceramic Society 90~(2)
  (2007) 337--357.

\bibitem{vasp_1}
G.~Kresse, J.~Hafner,
  \href{https://link.aps.org/doi/10.1103/PhysRevB.47.558}{Ab initio molecular
  dynamics for liquid metals}, Phys. Rev. B 47 (1993) 558--561.
\newblock \href {https://doi.org/10.1103/PhysRevB.47.558}
  {\path{doi:10.1103/PhysRevB.47.558}}.
\newline\urlprefix\url{https://link.aps.org/doi/10.1103/PhysRevB.47.558}

\bibitem{vasp_2}
G.~Kresse, J.~Furthm{\"u}ller, Efficient iterative schemes for ab initio
  total-energy calculations using a plane-wave basis set, Physical review B
  54~(16) (1996) 11169.

\bibitem{vasp_3}
G.~Kresse, D.~Joubert, From ultrasoft pseudopotentials to the projector
  augmented-wave method, Physical review b 59~(3) (1999) 1758.

\bibitem{monkhorst_pack}
H.~J. Monkhorst, J.~D. Pack, Special points for brillouin-zone integrations,
  Physical review B 13~(12) (1976) 5188.

\bibitem{methfessel_paxton}
M.~Methfessel, A.~T. Paxton,
  \href{https://link.aps.org/doi/10.1103/PhysRevB.40.3616}{High-precision
  sampling for brillouin-zone integration in metals}, Phys. Rev. B 40 (1989)
  3616--3621.
\newblock \href {https://doi.org/10.1103/PhysRevB.40.3616}
  {\path{doi:10.1103/PhysRevB.40.3616}}.
\newline\urlprefix\url{https://link.aps.org/doi/10.1103/PhysRevB.40.3616}

\bibitem{lammps}
S.~Plimpton, \href{http://lammps.sandia.gov.}{{Fast parallel algorithms for
  short-range molecular dynamics}}, Journal of computational physics 117~(1)
  (1995) 1--19.
\newline\urlprefix\url{http://lammps.sandia.gov.}

\bibitem{phonopy_1}
A.~Togo, I.~Tanaka, First principles phonon calculations in materials science,
  Scr. Mater. 108 (2015) 1--5.

\bibitem{phonopy_2}
A.~Togo, First-principles phonon calculations with phonopy and phono3py, J.
  Phys. Soc. Jpn. 92~(1) (2023) 012001.
\newblock \href {https://doi.org/10.7566/JPSJ.92.012001}
  {\path{doi:10.7566/JPSJ.92.012001}}.

\bibitem{exp_Mg_lc}
E.~R. Jette, F.~Foote, Precision determination of lattice constants, The
  Journal of Chemical Physics 3~(10) (1935) 605--616.

\bibitem{pei_2017_Mg}
Z.~Pei, X.~Zhang, T.~Hickel, M.~Fri{\'a}k, S.~Sandl{\"o}bes, B.~Dutta,
  J.~Neugebauer, Atomic structures of twin boundaries in hexagonal close-packed
  metallic crystals with particular focus on mg, NPJ Computational Materials
  3~(1) (2017) 1--7.

\bibitem{kim_Mg_Al_pot}
Y.-M. Kim, N.~J. Kim, B.-J. Lee, Atomistic modeling of pure mg and mg--al
  systems, Calphad 33~(4) (2009) 650--657.

\end{thebibliography}




\end{document}